%% file: main.tex
\documentclass[conference,a4paper]{IEEEtran}

\usepackage[utf8]{inputenc} 
\usepackage[T1]{fontenc}
\usepackage{url}              
\usepackage{cite}             

\usepackage[cmex10]{amsmath}  
\usepackage{mleftright}       
\mleftright                   

\usepackage{graphicx}        
\usepackage{booktabs}         
\usepackage{enumerate}
\usepackage{enumitem}
\usepackage{graphics}
\usepackage{amsthm}
\usepackage[compact]{titlesec}
\usepackage{setspace}
\usepackage{float}
\usepackage{caption}
\usepackage{subcaption}
\usepackage{balance} 
\usepackage{threeparttable}

\usepackage[cmintegrals]{newtxmath}
\usepackage{wrapfig}
\usepackage{tikz}
\usepackage{lipsum,adjustbox}
\usepackage{algpseudocode}
\usepackage{algorithm}
\usepackage{multirow}
\usepackage{fixmath}
\usepackage{comment}
\usepackage{stmaryrd}
\algnewcommand\algorithmicforeach{\textbf{for each}}
\algdef{S}[FOR]{ForEach}[1]{\algorithmicforeach\ #1\ \algorithmicdo}

\newcommand*\concat{\mathbin{\|}}
\newtheorem{theorem}{Theorem}[section]
\theoremstyle{definition}
\newtheorem{definition}{Definition}

\newcommand{\ms}[1]{\mathsf{#1}}
\newcommand{\mc}[1]{\mathcal{#1}}
\newcommand{\Set}[1]{\{#1\}}
\newcommand{\secp}{\lambda}

\newcommand{\mife}{\ms{MIFE}}
\newcommand{\sife}{\ms{SIFE}}
\newcommand{\qmife}{\ms{qMIFE}}

\newcommand{\msk}{\ms{MSK}}
\newcommand{\pp}{\ms{PP}}
\newcommand{\ct}{\ms{CT}}

\newcommand{\ek}{\ms{EK}}
\newcommand{\sk}{\ms{SK}}

\newcommand{\setup}{\ms{Setup}}
\newcommand{\KeyGen}{\mathsf{KeyGen}}
\newcommand{\enc}{\ms{Enc}}
\newcommand{\dec}{\ms{Dec}}

\newcommand{\dada}{, \ldots ,}
\newcommand{\mcF}{\mc{F}}
\newcommand{\mcX}{\mc{X}}
\newcommand{\mcY}{\mc{Y}}

\newcommand{\adv}{\mc{A}}
\newcommand{\negl}{\ms{negl}}

\newcommand{\mifeCS}{\mc{CS}}
\newcommand{\mifeMS}{\mc{MS}}
\newcommand{\mifeFS}{\mc{FS}}

\newcommand{\vecc}{\mathbold{c}}
\newcommand{\vecx}{\mathbold{x}}

\begin{document}
\bstctlcite{IEEEexample:BSTcontrol}
\title{ Quadratic Functional Encryption for Secure Training in Vertical Federated Learning} 


%

%
\author{%
  \IEEEauthorblockN{Shuangyi Chen\IEEEauthorrefmark{1},
                    Anuja Modi\IEEEauthorrefmark{2},
                    Shweta Agrawal\IEEEauthorrefmark{2},
                    and Ashish Khisti\IEEEauthorrefmark{1}}
  \IEEEauthorblockA{\IEEEauthorrefmark{1}%
                    University of Toronto,
                    \{shuangyi.chen@mail.utoronto.ca, akhisti@ece.utoronto.ca\}}
  \IEEEauthorblockA{\IEEEauthorrefmark{2}%
                    IIT Madras,
                    \{cs21d405@cse.iitm.ac.in, shweta.a@cse.iitm.ac.in\}}
}

\maketitle
\thispagestyle{plain}
\pagestyle{plain}
\begin{abstract}
  Vertical federated learning (VFL) enables the collaborative training of machine learning (ML) models in settings where the data is distributed amongst multiple parties who wish to protect the privacy of their individual data. Notably, in VFL, the labels are available to a single party and the complete feature set is formed only when data from all parties is combined. Recently, Xu et al. \cite{FedV} proposed a new framework called {\it FedV} for secure gradient computation for VFL using multi-input functional encryption. In this work, we explain how some of the information leakage in Xu et al.\ can be avoided by using Quadratic functional encryption when training generalized linear models for vertical federated learning.
\end{abstract}

\input{intro.tex}
\input{model}
\input{prelims}

\input{protocol.tex}
\input{conclusion.tex}



\bibliographystyle{IEEEtran}
\bibliography{main}

\balance

\clearpage
\appendices
\input{app-prelims.tex}
\input{pseudocode.tex}
\input{nonlinear.tex}
\end{document}

%% file: intro.tex
\section{Introduction}
\label{sec:intro}
In many emerging applications, a machine learning (ML) model must be trained using private data that is distributed among multiple parties. We study the setting of {\it vertical federated learning} (VFL) where each individual party has access to a subset of features and labels and must cooperate to train a ML model that makes use of all the features. When privacy of user data is required, homomorphic encryption (HE) \cite{paillier1999public,acar2018survey}, which enables the computation on encrypted data, provides a natural solution. In recent years, there has been a significant interest in HE based VFL systems, see e.g.,~\cite{PER,Yang2019AQM,yang2019parallel, 10.1145/3523150.3523171, he2021secure,10.1145/3529836.3529841,10.1155/2022/5094830,9101628}. 
Some works such as~\cite{yang2019parallel, 10.1145/3523150.3523171,10.1155/2022/5094830} consider a two-party protocol without the  trusted coordinator, while others~\cite{10.1145/3529836.3529841,he2021secure}.  consider a multi-party settings. Those frameworks require a large amount
of peer-to-peer communications. References \cite{PER,Yang2019AQM} propose frameworks comprised of one trusted coordinator, storing the global weights, and two parties, each with a subset of vertically partitioned data. However, these frameworks require the trusted coordinator to share plaintext global weights with parties, which undermines the model's confidentiality.

In a recent work, Xu et. al. \cite{FedV} proposed a generic and efficient privacy-preserving vertical Federated Learning (VFL) framework known as \textit{FedV} in the multiparty setting.  \textit{FedV} makes use of {\em single-input function encryption} ($\sife$) and {\em multi-input function encryption} ($\mife$), and makes the communication between the clients and the aggregator a one-round interaction. However, \textit{FedV} 
still have some key drawbacks. The protocol can reveal more information  to the aggregator than just the final gradient in each iteration. Moreover, the protocol reveals the respective updated weights in each iteration to clients, which additionally creates leakage. For more details, please see Section \ref{sec:protocol}. 

\subsection{Our Results.}
We observe that the leakage created in \textit{FedV} is caused by choosing an multi-input functional encryption ($\mife$) scheme that only supports linear functions. Due to this, the weights are required to be provided to each party for inclusion in encryption, which creates unnecessary leakage. We observe that for  linear models, this leakage can be prevented by using a more powerful $\mife$ scheme, namely $\mife$ for {\it quadratic functions} which can also be constructed using standard assumptions in cryptography \cite{AGT21, AGT22-TCC22}. As our main contribution in this work, we demonstrate how such a function encryption scheme can be applied in VFL training by proposing a novel construction of function vectors that serve as a basis for generating decryption keys. Our approach leads to direct computation of the gradients, without leakage of any intermediate results as is the case with \textit{FedV}. We discuss our proposed protocol, \textit{SFedV}, for linear model training in Section~\ref{sec:protocol} and the extension to logistic regression model in Appendix~\ref{ext}. We provide a thorough analysis of both security and efficiency in Section~\ref{sec:protocol}.  

%% file: model.tex
\section{System Model}
\label{sec:system}

\subsection{System Overview}
\label{sec:overview}
Our system model involves three types of entities: aggregator, $N$ clients, and Trusted Third Party (TTP). In the $t$th iteration for $t \in [T]$, each client holds a subset of features $\mathbold{X}_{i}^t \in \mathbb{R}^{S \times F_i }$ where $F_i$ is the number of features that client party $i$ holds and $S$ is the batch size. A complete feature set of the current iteration is expressed as $\mathbold{X}^t = [\mathbold{X}_0^t\mathbin\Vert...\mathbin\Vert\mathbold{X}_{N-1}^t] \in \mathbb{R}^{S \times F}$. One of the client parties holds the corresponding labels $\mathbold{y}^t \in \mathbb{R}^{S \times 1}$. The aggregator holds the entire model weights $\mathbold{w}^t=[\mathbold{w}^t_0\mathbin\Vert\mathbold{w}^t_1\mathbin\Vert...\concat\mathbold{w}^t_{N-1}] \in \mathbb{R}^{F \times 1}$ where $\mathbold{w}^t_i\in \mathbb{R}^{F_i \times 1}$ is the partial weights that pertains to $\mathbold{X}_i^t$. The aggregator is responsible for computing the model weights and the TTP is responsible for the generation of keys.  In this work we focus on linear models of the form: $f(\mathbold{X}^t,\mathbold{w}^t)=\mathbold{X}^t\cdot\mathbold{w}^t$  with a squared-error loss function: $L(\mathbold{w}^t) = \frac{1}{S}\sum_{s\in [S]} || \mathbold{y}^t [s]- (\mathbold{X}^t\cdot\mathbold{w}^t)[s] ||^2$. In our discussion, we will define the prediction error as:
{\small
\begin{equation}
\mathbold{u}^t = (\mathbold{y}^t-\mathbold{X}_{0}^t\cdot\mathbold{w}_{0}^t-...-\mathbold{X}_{N-1}^t\cdot\mathbold{w}_{N-1}^t).\label{u-error}
\end{equation}}
The gradient of $L(\mathbold{w}^t)$ with respect to $\mathbold{w}^t$ is expressed as
{\small
\begin{equation}
\begin{aligned}
g(\mathbold{w}^t)= -\frac{2}{S}
    \begin{bmatrix}
    \mathbold{y^t}^\top\mathbold{X}_{0}^t-\sum^{N-1}_{i=0}{\mathbold{w}_{i}^t}^\top{\mathbold{X}_{i}^t}^\top\mathbold{X}_{0}^t
    \\
    ...\mathbin\Vert\\
    {\mathbold{y}^t}^\top\mathbold{X}_{N-1}^t-\sum^{N-1}_{i=0}{\mathbold{w}_{i}^t}^\top{\mathbold{X}_{i}^t}^\top\mathbold{X}_{N-1}^t
    \end{bmatrix} \in \mathbb{R}^{1\times F} \label{q-goal}
\end{aligned}
\end{equation}}
The gradient is used to update the global weights in each iteration according to $\mathbold{w}^{t+1} =\mathbold{w}^t-\alpha g(\mathbold{w}^t)$ where $\alpha$ is the learning rate. We also discuss logistic regression model in Appendix~\ref{ext}. In each iteration of the training phase, our protocol takes as input an encrypted copy of the features $\mathbold{X}_{i}^t$ and encrypted labels $\mathbold{y}^t \in \mathbb{R}^{S}$ from clients, and collaboratively and securely computes the gradients $g(\mathbold{w}^t)$. 

Our threat model is defined as follows: we assume the aggregator is honest-but-curious meaning it correctly follows the algorithms and protocols but will try to infer clients' private data. Additionally, we assume that the aggregator does not collude with anyone. Similarly, the trusted third party is assumed not to collude with anyone. With respect to the clients, we assume that there are at most $N-1$ dishonest clients who may collude together and share their data to infer honest clients' information. 




The protocol enables all the entities to collaboratively compute the gradient using vertically partitioned data. During the training process, we aim to achieve the following privacy requirements: 1) The client $i$ and the aggregator should learn nothing about data $\mathbold{X}_j$ of client $j$ for $i\ne j$. 2) Any client should learn nothing about the trained global model weights $\mathbold{w}$, intermediate results including the prediction error as in \eqref{u-error} and the gradient $g(\mathbold{w})$. Moreover, $i$th client should not learn anything about his/her own corresponding weights $\mathbold{w}_{i}$. 


%% file: prelims.tex
\section{Preliminaries}
\label{sec:Preliminaries}

\subsection{Functional Encryption}
Functional Encryption \cite{SW05, BW07, BSW11} is a public key encryption scheme that enables fine-grained access control over the encrypted data. In Single Input Functional Encryption($\sife$), the secret key is associated with a function $f$, and the ciphertext is associated with the vector $\vecx$. The decryption of ciphertext using the secret key outputs $f(\vecx)$. Intuitively, the security says that the adversary learns nothing about the input $\vecx$ beyond what is revealed by $\{f_i(\vecx)\}_i$ for any set of secret keys corresponding to the functions $\{f_i\}_i$ that the adversary holds.
\subsection{Multi-Input Functional Encryption}
Goldwasser et al. \cite{GGG+14} generalized the functional encryption to support functions with multiple inputs. Multi-Input Functional Encryption, denoted as $\mife$ supports functions with arity greater than one. In $\mife$, the secret key is associated with a function $f$, and the $i$th  ciphertext is associated with the vector $\vecx_i$ for $i\in [N]$ where $N$ is the arity of the function $f$. The decryption of all the ciphertexts using the secret key outputs $f(\vecx_1, \ldots, \vecx_N)$. We now describe this notion in more detail.

\begin{definition}[Multi-Input Functional Encryption ($\mife$) \cite{AGT22-TCC22}]
\label{defi-MIFE}
\label{def:mcfe} 


\textbf{Syntax.} Let $N$ be the number of encryption slots, and $\mcF = \{\mcF_N\}_{N \in \mathbb{N}}$ be a function family such that, for all $f \in \mcF_N$, $f: \mcX_{1} \times \cdots \times \mcX_{N} \to \mcY$. Here $\mcX_i$ and $\mcY$ be the input and output spaces (respectively). A multi-input functional encryption $(\mife)$ scheme for function family $\mcF$ consists of the following algorithms.

\begin{description}
\item [$\ms{Setup}(1^{\secp}, 1^N) \rightarrow (\pp, \{\ek_i\}_{i}, \msk)$.] It takes a security parameter $1^{\secp}$, number of slots $1^N$, and outputs a public parameter $\pp$, $N$ encryption keys $\{\ek_i\}_{i \in [N]}$ and a master secret key $\msk$. (The remaining algorithms
implicitly take $\pp$ as input.)

\item [$\enc(\ek_i, \vecx) \rightarrow \ct_i$.] It takes the $i$th encryption key $\ek_i$ and an input $\vecx \in \mcX_{i}$, and outputs a ciphertext $\ct_{i}$.

\item [$\KeyGen( \msk, f ) \rightarrow \sk$.] It takes the master secret key $\msk$ and function $f \in \mcF$ as inputs, and outputs a secret key $\sk$.

\item [$\dec( \ct_{1} \dada \ct_{N}, \sk) \rightarrow y$.] It takes $n$ ciphertexts $ \ct_{1} \dada \ct_{N}$ and secret key $\sk$, and outputs a decryption value $y \in \mcY$ or a special abort symbol $\bot$.
\end{description}

\noindent{\textbf{Correctness.}}
An $\mife$ scheme for the function family $\mcF$ is correct if for all $\secp, N \in \mathbb{N},\; (\vecx_{1} \dada \vecx_{N}) \in \mcX_{1} \times \cdots \times \mcX_{N},\; f \in \mcF_N$, we have
{
\footnotesize{
\begin{align*}
\Pr\left[y =f(x_{1} \dada x_{N}) : 
\begin{array}{l}
(\pp, \{\ek_i\}_{i}, \msk) \gets \setup(1^{\secp}, 1^N)\\
\Set{\ct_{i} \gets \enc(\ek_i, \vecx)}_{i \in [N]} \\
\sk \gets \KeyGen( \msk, f)\\
y = \dec( \ct_{1}\dada \ct_{N}, \sk)
\end{array}
\right]
= 1.
\end{align*}}}
\end{definition}

\noindent{\textbf{Security.}} Intuitively, security says that no information about the messages can be learned by the adversary except what is revealed by virtue of functionality -- in more detail, an adversary possessing some ciphertexts and secret keys can perform decryption and learn the output of the functionality, which itself leaks something about the underlying plaintext. But besides this necessary leakage, the adversary does not learn anything. We provide the formal definition of security in Appendix~\ref{app:mife}.

 \noindent{\textbf{Multi-Input FE for Quadratic Functions.}} Agrawal, Goyal, and Tomida \cite{AGT22-TCC22} constructed a multi-input functional encryption for quadratic functions ($\qmife$). Let us define the $N$ input quadratic function $f$ as $f(\vecx_1,\ldots, \vecx_N) = \langle \vecc, \vecx \otimes \vecx \rangle$ where $\vecx = (\vecx_1||\ldots||\vecx_N)$. Here $\otimes$ denotes the Kronecker product. A $n$-input $\mife$ scheme for the function class $\mcF_{m, n}$ is defined as: each $i$th client encrypts $\vecx_i \in \mathbb{Z}^m$ using $i$th encryption key $\ek_i$ to get the $i$th ciphertext $\ct_i$ for $i \in [n]$. The $\KeyGen$ algorithm issues the secret key $\sk$ for $\vecc \in \mathbb{Z}^{(mn)^2}$ where $\vecc$ is the vector representation of the function $f \in \mcF_{m, n}$. The $\dec$ algorithm uses the secret key $\sk$ to decrypt $\ct_1, \ldots, \ct_n$ to get $\langle \vecc, \vecx \otimes \vecx \rangle$ and nothing else.

\subsection{FedV} 
As the system model of \textit{SFedV} (Section~\ref{sec:overview}), \textit{FedV} involves an aggregator, N clients, and a Trusted Third Party (TTP). Each client holds a subset of features $\mathbold{X}_{i}\in \mathbb{R}^{S \times F_i}$, and the first client also has the corresponding labels $\mathbold{y}\in\mathbb{R}^{S}$ along with its subset of features. The aggregator holds the complete model weights $\mathbold{w}=[\mathbold{w}_0||\mathbold{w}_1||...||\mathbold{w}_{N-1}]$ where $\mathbold{w}_{i}\in \mathbb{R}^{F_i}$ is the partial weights that pertains to $\mathbold{X}_{i}$. In each iteration, there are two steps to compute the gradient 
{\small
\begin{align}
    g(\mathbold{w})=-\frac{2}{S}\begin{bmatrix}
\mathbold{u}^\top\mathbold{X}_{0}\mathbin\Vert...\mathbin\Vert\mathbold{u}^\top\mathbold{X}_{N-1} \label{gradient_u}
\end{bmatrix}
\end{align}}
where $(\mathbold{u})^\top = (\mathbold{y}-\mathbold{X}_{0}\mathbold{w}_{0}-...-\mathbold{X}_{N-1}\mathbold{w}_{N-1})^\top$. 

In the first step called Feature Dimension Secure Aggregation, \textit{FedV} uses a Multi-Input Functional Encryption $(\mife)$ scheme for the inner product functionality \cite{ABG19, Libert} to securely compute the prediction error $\mathbold{u}$. In this step, the aggregator sends each $i$th client the partial weights $\mathbold{w}_{i}$. Then each $i$th client encrypts each sample of $(-\mathbold{X}_i  \mathbold{w}_{i}) \in \mathbb{R}^{S}$ and sends the ciphertext set $\ct_{-\mathbold{X}_i  \mathbold{w}_{i}}^{\mife}$ to the aggregator. The first client, holding the label $\mathbold{y}$, encrypts each sample of $\mathbold{y} - \mathbold{X}_1  \mathbold{w}_{1}$ and sends the ciphertext set $\ct_{\mathbold{y} - \mathbold{X}_1  \mathbold{w}_{1}}^{\mife}$ to the aggregator. The aggregator asks the TTP for the secret key $\sk_{\mathbold{v}}^{\mife}$ corresponding to the fusion vector $\mathbold{v}$. This vector $\mathbold{v}$ can be a binary vector where one in $i$th position means that the aggregator has received ciphertext from client $i$. Using the secret key $\sk_{\mathbold{v}}^{\mife}$, the aggregator decrypts the ciphertexts $\{\{\ct_{-\mathbold{X}_i  \mathbold{w}_{i}}^{\mife}\}_{i=1}^{N-1},\ct_{\mathbold{y} - \mathbold{X}_1  \mathbold{w}_{1}}^{\mife}\}$
to get the prediction error $\mathbold{u}$ (Equation~\eqref{u-error}), which is the inner product of the fusion vector $\mathbold{v}$ and the partial predictions from clients. 

In the second step called Sample Dimension Secure Aggregation, \textit{FedV} uses Single-Input Functional Encryption $(\sife)$ scheme to compute the gradient $g(\mathbold{w})$. In this step, each client $i$ encrypts each element of $\mathbold{X}_i$ and sends the ciphertext set $\ct_{\mathbold{X}_i}^{\sife}$ to the aggregator. On receiving the secret key $\sk_{\mathbold{u}}^{\sife}$ corresponding to the prediction error $\mathbold{u}$ from TTP, the aggregator decrypts the ciphertexts to get $\{\mathbold{u}^\top\mathbold{X}_{i}\}_{i\in[N]}$. The aggregator further processes the decryption results $\{\mathbold{u}^\top\mathbold{X}_{i}\}_{i\in[N]}$ according to Equation~\eqref{gradient_u} to get the gradient $g(\mathbold{w})$. Using the gradients, the model weights are updated and then the training of the next epoch starts. Note the transmission of $\mife$ ciphertext ($\ct_{\mathbold{X}_i  \mathbold{w}_{i}}^{\mife}$ or $\ct_{\mathbold{y} - \mathbold{X}_1  \mathbold{w}_{1}}^{\mife}$) and $\sife$ ciphertext ($\ct_{\mathbold{X}_i}^{\sife}$) can be simultaneous. Thus the communication between each client and the aggregator is a one-round interaction in each iteration.

\textbf{Leakage in FedV.} While \textit{FedV} preserves each client's data, it reveals the intermediate result, the prediction error $\mathbold{u}$ to the aggregator. Moreover, in the Feature Dimension Secure Aggregation step, the  $i$th client is required to know the respective weight $\mathbold{w}_{i}$. Additionally, the aggregator can use the secret key of $t$th iteration to decrypt the ciphertext of some other iteration $t'$ for $t' \ne t$ to infer more information about client data.

%% file: protocol.tex
\section{The Protocol}
\label{sec:protocol}
Now we introduce our protocol with Multi-input Quadratic Functional Encryption $\qmife$ \cite{AGT22-TCC22} as a privacy enhancement technology to do training in VFL setting.

At the beginning of the training phase, the aggregator initializes the global weights $\mathbold{w}^0$ and starts training. The training phase is iterative, where in the $t$th iteration, the TTP runs the $\qmife.\setup$ algorithm to get the public parameters $\pp^t$, $N$ encryption keys $\{\ek_i\}_{i \in [N]}^t$ and a master secret key $\msk^t$, then delivers the encryption key $\textsf{EK}_{i}^t$ to the corresponding client $i$.
After receiving the encryption key and determining the batch $\mathbold{X}^t$ used for training in this iteration, each client uses $\textsf{EK}_{i}^t$ to encrypt $\mathbold{x}_{i}^t$ which is the vectorized ${\mathbold{X}_{i}^t}$ (\textsf{vec}$(\cdot)$ stacks the columns of a matrix into a vector) to get ciphertext $\textsf{CT}_{i}^t$. Each client sends $\textsf{CT}_{i}^t$ to the aggregator. The client that holds the labels encrypts $\mathbold{x}_{i}^t$ and $\mathbold{y}^t$ with $\textsf{EK}_{i}^t$ to get ciphertexts $\textsf{CT}_{i}^t$ and $\textsf{CT}_{\mathbold{y}}^t$ respectively and then sends $(\textsf{CT}_{i}^t,\textsf{CT}_{\mathbold{y}}^t)$ to the aggregator. At the same time, the aggregator computes a set of function vectors $\mathcal{C}^t$ according to the model weights $\mathbold{w}^t$ of the current iteration and sends them to TTP to generate a set of decryption keys. Detailed procedure is described in Section~\ref{c-construction}. Then, the aggregator decrypts all the ciphertexts $(\{\textsf{CT}_i^t\}_{i=0}^{N-1},\textsf{CT}_y^t)$ that were received from clients using the secret keys received from TTP, to get each element of \eqref{q-goal} respectively. By concatenating those elements and further processing the results, the aggregator gets the gradients. After this, it can update the global weights and start the training of the next iteration. Algorithm~\ref{alg:training} shows the training procedure for linear models and also supports the training for logistic regression as discussed in Appendix~\ref{ext}.

\begin{tiny}
     \begin{algorithm}
\caption{Training Procedure} 
\label{alg:training}
\begin{algorithmic}[1]
\Procedure{Training-Aggregator}{$\mathbold{w}^t,s,\{F_i\}_{i=0}^{N-1},N$}
    \State $\mathbold{res} = \textbf{0}^{F}$
    \ForEach{$i \in {0,...,N-1}$}
    \ForEach{$p \in {0,...,F_{i}-1}$}
    \State $\mathbold{c}_{i,p}^t :=$ CGEN($\mathbold{w}^t,S,{F},N,i,p$)
    \EndFor
    \EndFor
    \State $\mathcal{C}^t:=\{\mathbold{c}^t_{i,p}, i\in [N],p \in [F_{i}]\}$
    \State $\{\{\qmife.\sk^t_{\mathbold{c}_{i, p}}\}_{p=0}^{F_i}\}_{i=0}^{N-1}= \textsf{obtain-dk-from-TTP}(\mathcal{C}^t)$
    \ForEach{$i \in {0,...,N-1}$}
    \If{party $i$ has label $\mathbold{y}^t$} 
    \State ($\ct_i^t,\ct_y^t$)= \textsf{obtain-ct-from-client()}
    \Else
    \State $\ct_i^t$ = \textsf{obtain-ct-from-client()}
    \EndIf
    \EndFor
    \State ${\ct}^t =\{\{{\ct}^t_{i}\}_{i=0}^{N-1}, {\ct}^t_{y}\}$ 
    \ForEach{$n \in {0,...,N-1}$}
    \ForEach{$p \in {0,...,F_{n}-1}$}
        \State \textsf{idx} =  $\sum_{i=0}^{n-1}F_{i}+p$
        \State $\mathbold{res}[\textsf{idx}] =  \qmife.\dec( {\ct}^t, \qmife.\sk^t_{\mathbold{c}_{n,p}})$
    \EndFor
    \EndFor
    \State $\nabla L(\mathbold{w}^t) = -\frac{2}{S}\mathbold{res} + \lambda \nabla R(\mathbold{w}^t)$
    \State $\mathbold{w}^{t+1} = \mathbold{w}^t - \alpha \nabla L(\mathbold{w}^t)$
    \EndProcedure
\Procedure{Training-Client}{$ \mathbold{X}_{i}^t$}
\Function{obtain-ct-from-client()}{}
    \State $\qmife.\ek_{i}^t$= obtain-ek-from-TTP()
    \State $\mathbold{x}_{i}^t:= \textsf{vec}({\mathbold{X}_{i}^t})$ 
    \If{party $i$ has label $\mathbold{y}^t$} 
    \State $\ct_{i}^t:= \qmife.\enc(\qmife.\ek^t_{i}, \mathbold{x}_{i}^t)$
    \State  $\ct_{y}^t:= \qmife.\enc(\qmife.\ek_{i}^t, \mathbold{y}^t)$
    \State Return $(\ct^t_{i},\ct_{y}^t)$ to Aggregator 
    \Else 
    \State $\ct_{i}^t:= \qmife.\enc(\qmife.\ek_{i}^t, \mathbold{x}_{i}^t)$
    \State Return $\ct^t_{i}$ to Aggregator
    \EndIf
    \EndFunction
    \EndProcedure
\Procedure{Training-TTP}{$1^\secp, 1^N$}
    \Function{obtain-ek-from-TTP()}{}
    
    \State $(\pp^t, \{\ek_i\}^t_i, \msk^t) \gets \qmife.\setup(1^\secp, 1^N)$
    \State Deliver $\qmife.\ek^t_{i}$ to party $i$, $i\in[N]$
    \EndFunction
    \Function{obtain-dk-from-TTP}{$\mathcal{C}^t$}
    
    \ForEach{$i \in {0,...,N-1}$}
    \ForEach{$p \in {0,...,F_i}$}
    \State $\qmife.\KeyGen(\qmife.\msk^t, \mathbold{c}_{i,p}^t)\rightarrow \qmife.\sk^t_{\mathbold{c}_{i,p}}$
    \EndFor
    \EndFor
    \State Return $\{\{\qmife.\sk^t_{\mathbold{c}_{i,p}}\}_{p=0}^{F_i}\}_{i=0}^{N-1}$
    \EndFunction
    \EndProcedure
    \end{algorithmic}
\end{algorithm}
\end{tiny}

\subsection{Construction of Function Vectors} \label{c-construction}

Our goal is to compute the gradient $g({\mathbold{w})}$ (Equation \eqref{q-goal}).  For simplicity, we drop the superscript $t$ in our discussion. We define $\mathbold{x}=[\mathbold{x}_{0}||...||\mathbold{x}_{N-1}||\mathbold{y}]$ where $\mathbold{y}$ is the label vector and $\mathbold{x}_{i}$ is the vectorized $\mathbold{X}_{i}$. Recall $g(\mathbold{w})$ is a vector of length $F$, where $F$ is the total number of features. The key insight is that for the $f^{\text{th}},f\in[F]$ element in \eqref{q-goal}, we construct a function vector $\mathbold{c}_f$ based on weights $\mathbold{w}$ of current iteration, such that $g(\mathbold{w})[f]=-\frac{2}{S}\left\langle\mathbold{c}_f,\mathbold{x}\otimes\mathbold{x}\right\rangle$. Then the aggregator concatenates $g(\mathbold{w})[f], f\in [F]$ to obtain the gradients.

\begin{figure}
    \centering
\includegraphics[scale=0.31]{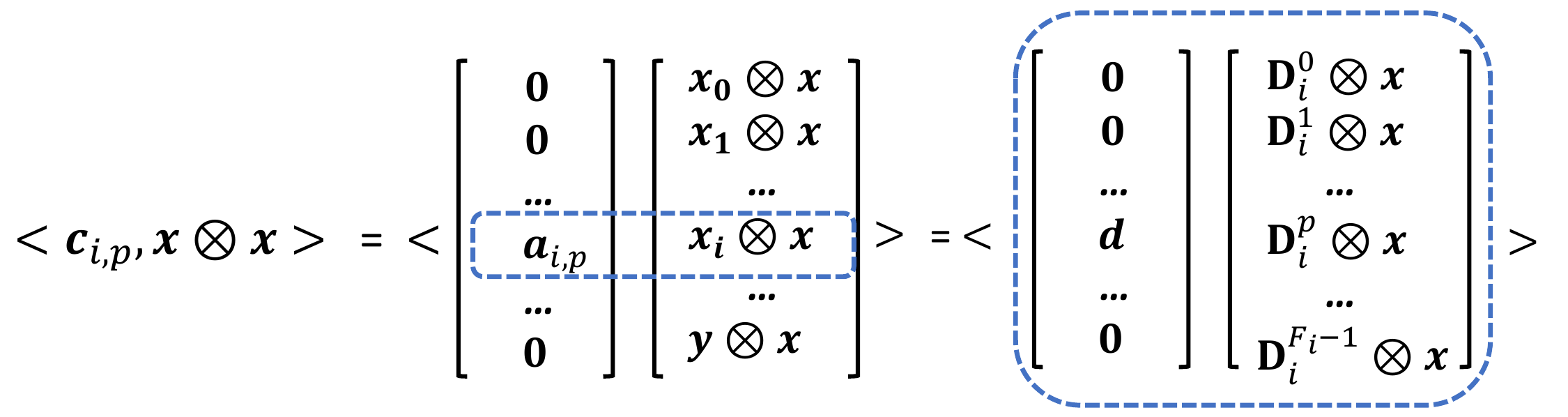} 
    \caption{Decomposition of $\langle \mathbold{c}_{i,p},\mathbold{x}\otimes\mathbold{x} \rangle$}
    \label{fig:decomp0}
\end{figure}

For simplicity, we define the following:
{\small{
\begin{align}
\mathbold{z}_i=\mathbold{u}^\top\mathbold{X}_{i},~~\mathbold{b}_{j}^i=\mathbold{w}_{j}^\top\mathbold{X}_{j}^\top\mathbold{X}_{i},~~\mathbold{b}_{y}^i=\mathbold{y}^\top\mathbold{X}_{i}.
\label{eq:b-def}
\end{align}}}
Now we decompose $g(\mathbold{w})$ and $\mathbold{x}\otimes\mathbold{x}$ to reduce the assignment. Note that to compute $g(\mathbold{w})$ it suffices to compute $\mathbold{z}_0,...,\mathbold{z}_{N-1}$ as in \eqref{gradient_u}.  Here we define a set of function vectors $\mathcal{C} = \{\mathcal{C}_{i}\}_{i=0}^{N-1}$ and $\mathcal{C}_{i} = \{\mathbold{c}_{i,p}\}_{p=0}^{F_i-1}$, where $\mathcal{C}_{i}$ is a subset of function vectors that are used to compute elements in $\mathbold{z}_i$, $\mathbold{c}_{i,p}$ is the function vector to compute $p$th element of $\mathbold{z}_i$ as in \eqref{eq6}. 
{\small{
\begin{eqnarray}
\mathbold{z}_i[p] = \langle \mathbold{c}_{i,p}, \mathbold{x}\otimes\mathbold{x} \rangle \label{eq6}
\end{eqnarray}}}
We construct $\mathbold{c}_{i,p}$ block by block according to the decomposition of $\mathbold{x}\otimes\mathbold{x} $. Consider dividing $\mathbold{x}\otimes\mathbold{x}$ into $N+1$ blocks as in the middle of Figure~\ref{fig:decomp0}. 

Since in the computation of $\mathbold{z}_i$, only the component $\mathbold{x}_i\otimes\mathbold{x} $ is required, we set \textbf{0} vector of the corresponding lengths as the coefficients of the blocks $\{\mathbold{x}_j \otimes \mathbold{x}\}_{j=0}^{N}$ if $j\neq i$. We design $\mathbold{a}_{i,p}$ to make $\mathbold{z}_i[p] = \langle \mathbold{a}_{i,p},\mathbold{x}_i\otimes\mathbold{x}\rangle$.

Let $\textsf{D}^f_{i}$ denotes the $f$th column of $\mathbold{X}_{i}$. Thus $\mathbold{x}_i = [\textsf{D}_{i}^{0};...;\textsf{D}_{i}^{F_i-1}]$. Note that in the computation of  $\mathbold{z}_i[p]$ only the column $\textsf{D}_{i}^{p}$ is used, thus we set \textbf{0} vector as the coefficients of $\{\textsf{D}_{i}^{q}\otimes \mathbold{x}\}_{q=0}^{F_i-1}$ if $q\neq p$. Hence we can express
$\langle \mathbold{a}_{i,p},\mathbold{x}_i\otimes\mathbold{x}\rangle = \langle \mathbold{d},\textsf{D}_i^p\otimes\mathbold{x}\rangle$. We design $\mathbold{d}$ to achieve $\mathbold{z}_i[p]=\langle \mathbold{d},\textsf{D}_i^p\otimes\mathbold{x}  \rangle$.

From \eqref{q-goal}, \eqref{gradient_u} and \eqref{eq:b-def} note that we can express:
{\small{
\begin{eqnarray}
    \mathbold{z}_i[p] = \sum_{j=0}^{N-1}-\mathbold{b}_{j}^i[p]+\mathbold{b}_{y}^i[p] \label{eq5}
\end{eqnarray}}}
where we have $\mathbold{z}_i[p]=(\mathbold{u}^\top\mathbold{X}_{i})[p]$, $\mathbold{b}_j^i[p] = (\mathbold{w}_j^\top\mathbold{X}_j^\top\mathbold{X}_i)[p]$,$\mathbold{b}_y^i[p] = (\mathbold{y}^\top\mathbold{X}_i)[p]$. We expand $\mathbold{b}_j^i[p]$ by first performing multiplication term-by-term and then summing products as in \eqref{eq12}. Based on this equation, we determine the method for constructing the coefficients vector $\mathbold{d}$.
{\small{
\begin{align}
    \mathbold{b}_j^i[p] &= (\mathbold{w}_j^\top\mathbold{X}_j^\top\mathbold{X}_i)[p] = (\mathbold{w}_j^\top\mathbold{X}_j^\top)\textsf{D}_i^p\nonumber \\
    &= \left[\sum_{f=0}^{F_j-1}\mathbold{w}_{j}^\top[f]\textsf{D}_i^f[0]\mathbin\Vert...\mathbin\Vert\sum_{f=0}^{F_j-1}\mathbold{w}_{j}^\top[f]\textsf{D}_i^f[S-1]\right]\textsf{D}_i^p \nonumber\\
    &= \sum_{s=0}^{S-1} \sum_{f=0}^{F_j-1}\mathbold{w}_j[f]\textsf{D}_j^f[s]\textsf{D}_i^p[s] \label{eq12}
\end{align}}}

\begin{figure}
    \centering
    \includegraphics[scale=0.27
    ]{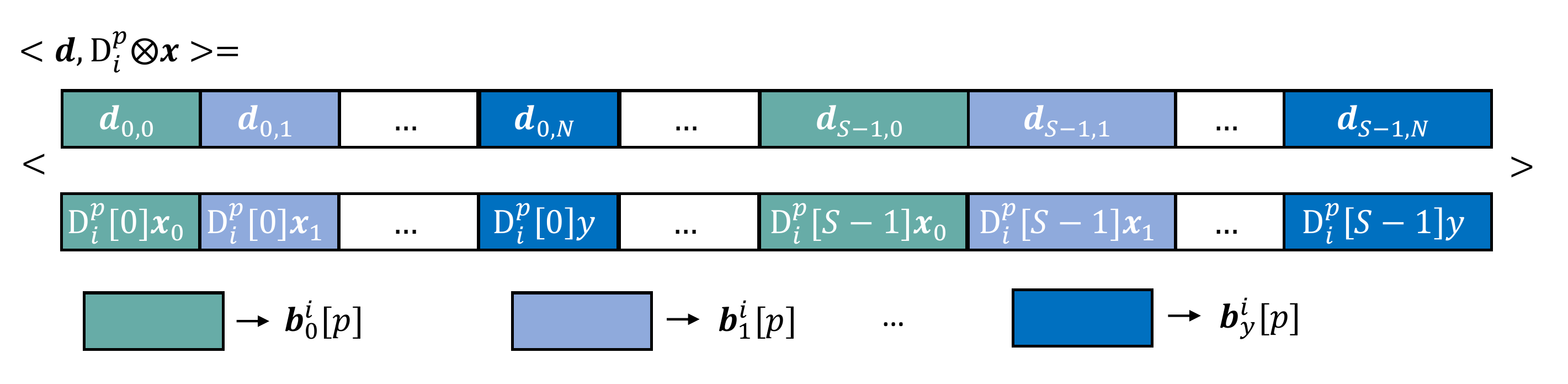} 
    \caption{Decomposition of $\langle \mathbold{d},\textsf{D}_i^p\otimes\mathbold{x} \rangle$}
    \label{fig:d-decomp}
\end{figure}

Now the goal is to construct $\mathbold{d}$ such that $\mathbold{z}_i[p] = \langle \mathbold{d},\textsf{D}_i^p\otimes\mathbold{x} \rangle$. 
We keep decomposing $\mathbold{d}$ and  $\textsf{D}_i^p\otimes\mathbold{x}$ to blocks as in Figure~\ref{fig:d-decomp}.
 \\
In Figure~\ref{fig:d-decomp}, blocks with the same color will be designed to compute the corresponding term on the right side of Equation \eqref{eq5}. Considering $\mathbold{d}_{s,j}, s\in[S]$, we can set the following relations:
{\small{
\begin{align}
    \mathbold{b}_j^i[p] = \sum_{s=0}^{S-1}\langle \mathbold{d}_{s,j}, \textsf{D}_i^p[s]\mathbold{x}_j \rangle\label{eq10}  \\
 \mathbold{b}_y^i[p] = \sum_{s=0}^{S-1}\langle \mathbold{d}_{s,N}, \textsf{D}_i^p[s]\mathbold{y} \rangle 
\end{align}}}
Next, we introduce the approach to construct $\mathbold{d}_{s,j}, j\in [N]$ according to the corresponding weight piece. We remove the outer summation in \eqref{eq12} and \eqref{eq10} to obtain:
{\small{
\begin{eqnarray}
    \sum_{f=0}^{F_j-1}\mathbold{w}_j[f]\textsf{D}_j^f[s]\textsf{D}_i^p[s] = \langle \mathbold{d}_{s,j},\textsf{D}_i^p[s]\mathbold{x}_j\rangle \label{eq13}
\end{eqnarray}}}

\begin{figure}[h]
    \centering
    \includegraphics[scale=0.24]{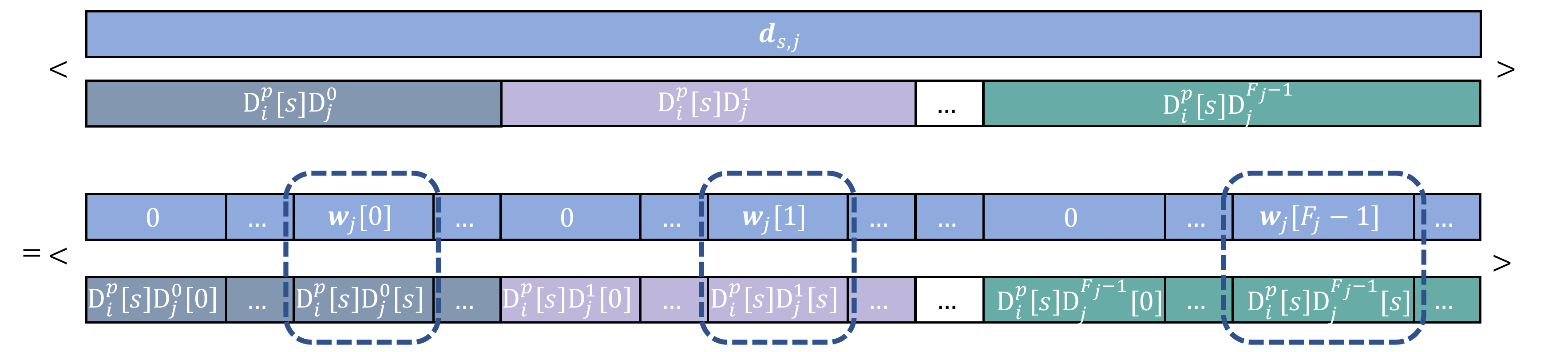} 
    \caption{Decomposition of $\langle \mathbold{d}_{s,j},\textsf{D}_i^p[s]\mathbold{x}_j\rangle$}
    \label{fig:decomp2}
\end{figure}

We design $\mathbold{d}_{s,j}$ to achieve \eqref{eq13} as Figure~\ref{fig:decomp2} shows. We decompose $\langle \mathbold{d}_{s,j},\textsf{D}_i^p[s]\mathbold{x}_j\rangle$ into blocks $\textsf{D}_i^p[s]\textsf{D}_j^f, f\in[F_j]$. In each block $\textsf{D}_i^p[s]\textsf{D}_j^f$, we take one entry $\textsf{D}_i^p[s]\textsf{D}_j^f[s]$ and set its coefficient to $\mathbold{w}_j[f]$ just as the left side of \eqref{eq13}. For other unneeded terms, we set the coefficient to \textbf{0}.

The approach to construct $\mathbold{d}_{s,j},j\in[N]$ can be easily extended to the case for $\mathbold{d}_{s,N}$. By designing the blocks of $\mathbold{d}$ this way, we can achieve $\mathbold{z}_i[p] = \langle \mathbold{d},\textsf{D}_i^p\otimes\mathbold{x} \rangle$. The algorithms to construct the function vectors $\mathbold{c}_{i,p}$ and $d$ are provided in Appendix~\ref{pseudocodealgo}.

\input{analysis}

%% file: analysis.tex
\subsection{Privacy Analysis}
Recall the aim of our framework. We want the client $i$ and the aggregator to learn nothing about data $\mathbold{X}_j$ of client $j$ for $i \ne j$. We also want that any client should learn nothing about the trained global model weights $\mathbold{w}$, intermediate results including error between labels and feed-forward output as in \eqref{u-error} and the gradient $g(\mathbold{w})$. Moreover, $i$th client should not learn anything about his/her own corresponding weights $\mathbold{w}_{i}$. 
In this section, we prove that we have achieved the above-stated goal.

\begin{theorem}
    If Quadratic MIFE $(\qmife)$ is secure according to definition \ref{def:mifesecurity}, then in each training iteration $t$, $i$th client's data $\mathbold{X}_i^t$ for $i \in [N]$ is hidden from client $j$ and the aggregator, trained global model weights $\mathbold{w}^t$ and intermediate results $\mathbold{u}^t$ as in \eqref{u-error} are hidden from the clients and $i$th client learns nothing about weight $\mathbold{w}_{i}$.
\end{theorem}

Let us fix the iteration number to be $t$. In each iteration, the TTP runs the $\qmife.\setup$ algorithm to get public parameters $\pp^t$, $N$ encryption keys $\{\ek_i\}_{i \in [N]}^t$ and a master secret key $\msk^t$. The clients encrypt their respective data and send the ciphertexts to the aggregator. The aggregator asks the TTP for the secret key corresponding to the set of vectors    $\mathcal{C}^t$. Each vector $\mathbold{c}_{i,p}^t$ is set in such a way that the $\qmife.\dec$ only reveals the inner product $\langle\mathbold{c}_{i,p}^t,\mathbold{x}\otimes\mathbold{x}\rangle = ((\mathbold{u}^t)^\top\mathbold{X}_{i}^t)[p]$. Quadratic MIFE ensures that nothing about $\mathbold{X}_{i}^t$ and $\mathbold{u}^t$ is revealed to the aggregator. Moreover, each client encrypts their data using different encryption keys. The ciphertexts are indistinguishable; hence, clients cannot predict other clients' data.

Unlike \textit{FedV}, in our framework, the client runs the $\qmife.\enc$ algorithm which only takes their respective data and encryption keys as input. Hence, each client $i$ learns nothing about their respective weight $\mathbold{w}_{i}^t$. Moreover, Quadratic MIFE ensures that client $i$ learns nothing about the global weight $\mathbold{w}^t$. The aggregator does not share the gradients in any form with the clients. Therefore, the gradient $g(\mathbold{w}^t)$ is also not revealed.

\textbf{Importance of using new $\qmife$ instance for each iteration.} Let in the iteration $t$, the ciphertext be $\ct^t$ and secret key be $\qmife.\sk^t$. Suppose the TTP uses the same $\msk$ to generate secret keys $\qmife.\sk^{t+1}$ for some other iteration, say $t+1$, then the aggregator may use $\qmife.\sk^{t+1}$ to decrypt the ciphertext  $\ct^t$ instead of using it to decrypt the ciphertext  $\ct^{t+1}$. Using this "mix-and-match" attack by performing decryptions of secret key and ciphertexts from different iterations, he will know $g(\mathbold{w})=-\frac{2}{S}\begin{bmatrix} (\mathbold{u}^{t+1})^\top\mathbold{X}_{0}^t||...||(\mathbold{u}^{t+1})^\top\mathbold{X}_{N-1}^t \end{bmatrix}$.

If TTP generates different $\qmife$ instance for every iteration, then decryption of $\ct^t$ with secret key $\qmife.\sk^{t+1}$ will give some garbage value which will be irrelevant for the aggregator. Therefore, it is important for the TTP to generate a new $\qmife$ instance for every iteration.

\textbf{Comparison of our framework with FedV.}  Unlike \textit{FedV}, our framework does not leak the intermediate result $\mathbold{u}$ to the aggregator. The global weights $\mathbold{w}$ are kept secret from the clients and each client also learns nothing about their respective weights. In addition to this, we also ensure that the aggregator cannot use the mix-and-match attack to learn some useful information. 

\subsection{Efficiency Analysis}


\noindent{\textbf{Communication.}} 
Regarding communication complexity, \textit{SFedV} requires one-way client-aggregator communication, while \textit{FedV} needs one-round client-aggregator communication due to the delivery of global weights by the aggregator. Additionally, \textit{SFedV} uses a new $\qmife$ instance in each iteration to prevent mix-and-match attacks. Thus, an increase of communication between TTP and clients becomes necessary. Note that \textit{FedV} can also prevent mix-and-match attacks by using new instances of $\mife$ and $\sife$ in each iteration. In such a scenario, the client-TTP communication complexity for each iteration will be the same for both \textit{FedV} and \textit{SFedV}. 


\begin{table}[h]
\caption{Comparison of \textit{FedV} and \textit{SFedV} regarding the number of encryption processes on each client and the number of decryption processes on the aggregator in each iteration.}
\label{tab:noencdec}
\centering
\begin{threeparttable}{\small
\begin{tabular}{cccc}
\toprule
                      & \multicolumn{2}{c}{\textit{FedV}} & \textit{SFedV} \\ 
                      & $\mife$       & $\sife$       & $\qmife$  \\ \midrule
Encryptions on each client & $S$           & $S\cdot F_i$          & $1$     \\
Decryptions on the aggregator & $S$           & $F$          & $F$      \\ \bottomrule
\end{tabular}}
\begin{tablenotes}
			\footnotesize
			\item $S$: Batch size. $F$: Total number of features. $F_i$: Number of features belonging to client $i$.
		\end{tablenotes}

\end{threeparttable} 
\end{table}  



\noindent{\textbf{Computation.}} 
Table~\ref{tab:noencdec} provides a comparison between \textit{FedV} and \textit{SFedV} in terms of the number of encryption and decryption processes in each iteration. The significant improvement of \textit{SFedV} is attributed to the advancement of quadratic $\mife$ and the careful design of function vectors.

In terms of the number of the vector corresponding to which the secret keys are generated, \textit{FedV} uses two vectors for two steps: $\mathbold{v}$ for feature dimension secure aggregation and $\mathbold{u}$ for sample dimension secure aggregation. In contrast, our \textit{SFedV} framework employs $F$ vectors $\mathbold{c}$, where $F$ is the total number of features. The increase in size can be justified by our use of a quadratic $\mife$ scheme instead of inner product $\mife$.

%% file: conclusion.tex
\section{Conclusions}
Prior $N$-party VFL framework \textit{FedV} incurs information leakage which seriously undermines individual data privacy. In this work, to address the privacy issues, We propose a leak-free protocol, called \textit{SFedV}, for multiparty VFL regression model training. Our approach simplifies the VFL pipeline and preserves the privacy of client data, model weights, and intermediate results, by designing special function vectors and using a quadratic MIFE scheme to compute gradients directly.

%% file: app-prelims.tex


\section{Security Definition for MIFE}
\label{app:mife}

In an indistinguishability-based security game between a challenger and an adversary, the challenger runs the $\setup$ algorithm to generate the public parameters $\pp$, $N$ encryption keys $\ek_i$, and master secret key $\msk$. The adversary then chooses the set of encryption keys that she wants. Then the adversary chooses two messages $\mathbold{x}^0$ and $\mathbold{x}^1$ and gives them to the challenger. The challenger chooses a bit $\beta$ at random and encrypts the message $\mathbold{x}^\beta$ using the $i^\text{th}$ encryption key $\ek_i$ to get challenge ciphertext $\ct_i$. The adversary then asks the challenger for the secret keys corresponding to the functions $f$. At last, the adversary guesses a bit $\beta'$ and replies to the challenger. The admissible adversary wins if $\beta' = \beta$. In security, we want the probability of the adversary winning the security game to be negligibly close to 1/2.

The adversary is said to be admissible if and only if she sends at least one element of the form $(i, *, *)$ in the message space and she queries the secret key for the function $f$ which satisfies the constraint that $f(\mathbold{x}^0) = f(\mathbold{x}^1)$. We formally define MIFE security in Definition  \ref{def:mifesecurity}.

\begin{definition}[MIFE Security \cite{AGT22-TCC22}] \label{def:mifesecurity}
An $\mife$ scheme is IND-secure if for any stateful \textit{admissible} PPT adversary $\adv$, there exists a negligible function $\negl(\cdot)$ such that for all $\secp, N \in \mathbb{N}$, the following probability is negligibly close to 1/2 in $\secp$:

\begin{align*}
\Pr\left[ \beta' = \beta : 
\begin{array}{l}
\beta \gets \{0,1\}\\
(\pp, \{\ek_i\}_{i \in [N]}, \msk) \gets\\
\setup(1^{\secp}, 1^N)\\
(\mifeCS, \mifeMS, \mifeFS) \gets \adv(1^\secp, \pp) \text{ s.t. } \\
\mifeCS \subseteq [N]\\
\mifeMS = \{i^\mu, \vecx^{\mu,0}, \vecx^{\mu, 1}\}_{\mu \in [q_c]}\\
\mifeFS = \{f^v\}_{v \in [q_k]}\\
\{\ct_{\mu} \gets \enc(\ek_{i^\mu}, \vecx^{\mu, \beta})\}_{\mu}\\
\{\sk_v \gets \KeyGen(\msk, f^v)\}_v\\
\beta' \gets \adv \left ( \{\ek_i\}_{i \in \mifeCS}, \{\ct_\mu\}_\mu, \{\sk_v\}_v\right )
\end{array}
\right]
\end{align*}
where the adversary $\adv$ is said to be admissible if and only if
\begin{itemize}
    \item $q_c[i]>0$ for all $i \in [N]$, where $q_c[i]$ denotes the number of elements of the form $(i, *, *)$ in $\mifeMS$.
    \item $f(\vecx_1^0, \ldots, \vecx_n^0) = f(\vecx_1^1, \ldots, \vecx_n^1)$ for all sequences $(\vecx_1^0, \ldots, \vecx_n^0, \vecx_1^1, \ldots, \vecx_n^1, f)$ such that:
    \begin{itemize}
        \item For all $i \in [n]$, $[(i, \vecx_i^0, \vecx_i^1) \in \mifeMS]$ or $[i \in \mifeCS \text{ and } \vecx_i^0 = \vecx_i^1]$,
        \item $f \in \mifeFS$.
    \end{itemize}
\end{itemize}
\end{definition}


    
    

%% file: pseudocode.tex
\section{Pseudocode}
\label{pseudocodealgo}

In this section, we give reference to the algorithms used for constructing the function vectors. The algorithm \ref{alg:makec} shows the construction of $\mathbold{c}_{i,p}$ and the algorithm \ref{alg:subc} shows the construction of vector $d$. 

\begin{algorithm}
\caption{Construct c} 
\label{alg:makec}
    \begin{algorithmic}[1]
      \Function{cgen}{$\mathbold{w},S,F,N,i,p$}
      \State $\mathcal{A} = \{\mathbold{a}_{i,n}\}_{n=0}^{N}$
      \ForEach{$n \in {0,...,N}$}
       \If{$n \neq i$ and $i\neq N$}
       \State $\mathbold{a}_{i,n} = \textbf{0}^{S^2 (F+1) F_i}$ 
       \ElsIf{$n == i$ and $i\neq N$}
       \State $\mathcal{D} = \{\mathbold{d}_{i}\}_{i=0}^{F_i-1}$
       \ForEach{$f \in {0,...,F_i-1}$}
       \If{$f == p$}
       \State $\mathbold{d}_{f} = d = $ SUBCGEN($\mathbold{w},S,F_{j},N$)
       \ElsIf{$f \neq p$}
       \State $\mathbold{d}_{f} = \textbf{0}^{S^{2}(F+1)}$
       \EndIf
       \State $\mathbold{a}_{i,n} = [\mathbold{d}_{0}; \ldots; \mathbold{d}_{F_i-1}]$
       \EndFor
       \ElsIf{$i==N$}
       \State $\mathbold{a}_{i,n} =\textbf{0}^{S^2 (F+1)}$
       \EndIf
      \EndFor
      \State Set $\mathbold{c}_{i,p} = [\mathbold{a}_{i,0};\ldots; \mathbold{a}_{i,N}] $
      \State \Return $\mathbold{c}_{i,p}$
      \EndFunction
    \end{algorithmic}
\end{algorithm}

\begin{algorithm}
\caption{Construct subc}
\label{alg:subc}
    \begin{algorithmic}[1]
      \Function{subcgen}{$\mathbold{w},S,F_{j},N$}
      \State $d = \{\mathbold{d}_{s,n}\}, s\in[S], n\in[N+1]$
      \ForEach{$s \in {0,...,S-1}$}
      \ForEach{$j \in {0,...,N}$}
      \If{$j\neq N$}
      \State $\mathbold{d}_{s,j} = \textbf{0}^{SF_j}$ \Comment{Initialization}
      \ForEach{$f \in {0,...,F_j-1}$}
      \State $\mathbold{d}_{s,j}[fS+s] = \mathbold{w}_j[f]$
      \EndFor
      \ElsIf{$j==N$}
      \State $\mathbold{d}_{s,N} = \textbf{0}^{S}$ \Comment{Initialization}
      \State $\mathbold{d}_{s,N}[s] = 1$
      \EndIf
      \EndFor
      \State $\mathbold{d}_{s} = [\mathbold{d}_{s,0};\ldots;\mathbold{d}_{s,N}]$
      \EndFor
        \State \Return $\mathbold{d} = [\mathbold{d}_{0};\ldots;\mathbold{d}_{S-1}]$
      \EndFunction
    \end{algorithmic}
\end{algorithm}

%% file: nonlinear.tex
\section{Extension to Logistic Regression Model} \label{ext}
In this section, we extend the protocol to work for logistic regression with the help of Taylor approximation. The prediction function of logistic models is as follows:
\begin{eqnarray}
f(\mathbold{x},\mathbold{w}) =  \frac{1}{1-e^{-\mathbold{x}\mathbold{w}}} 
\end{eqnarray}
\noindent We use Cross-Entropy as the loss function for logistic regression. The loss function in the vectorized form is

\begin{align}
\footnotesize
L(\mathbold{w}) = \frac{1}{s}\left[-\mathbold{y}^\top \log(f(\mathbold{X},\mathbold{w}))-(\textbf{1}-\mathbold{y})^\top \log(\textbf{1}-f(\mathbold{X},\mathbold{w}))\right]\label{crossentropy} \nonumber\\
\end{align}
\noindent Here we use Taylor approximation to make the loss function polynomial. In \cite{PER}, it takes a Taylor Series expansion of $\log(1+e^{-z})$ around $z=0$.
\begin{eqnarray}
\log(1+e^{-z}) = \log{2}-\frac{1}{2}z+\frac{1}{8}z^{2}-\frac{1}{192}z^4+O(z^6) \label{TaylorExpansion}
\end{eqnarray}
We apply \eqref{TaylorExpansion} to \eqref{crossentropy} and get the gradients of vector-format expression as given below.
\begin{eqnarray}
g(\mathbold{w}) \approx \frac{1}{S}\left(\frac{1}{4}\mathbold{X}\mathbold{w}-\mathbold{y}+\frac{1}{2}\right)^\top \mathbold{X} \label{vec-Taylor-CE-loss}
\end{eqnarray}
Then we decompose $\mathbold{X}\mathbold{w}= \mathbold{X}_{0}\mathbold{w}_{0}+\mathbold{X}_{1}\mathbold{w}_{1}+...+\mathbold{X}_{N-1}\mathbold{w}_{N-1}$ and $\mathbold{X}= [\mathbold{X}_{0}||\mathbold{X}_{1}||...||\mathbold{X}_{N-1}]$ and substitute the decomposition into (\ref{vec-Taylor-CE-loss}) to get
\begin{equation}
\begin{aligned}
&g(\mathbold{w})\approx  
&\frac{1}{S}
\begin{bmatrix}
    -(\mathbold{y}-\frac{1}{2})^\top\mathbold{X}_{0}+\frac{1}{4}\sum_{j=0}^{N-1}\mathbold{w}_{j}^\top\mathbold{X}_{j}^\top\mathbold{X}_{0}||\\
    ...||\\
    -(\mathbold{y}-\frac{1}{2})^\top\mathbold{X}_{N-1}+\frac{1}{4}\sum_{j=0}^{N-1}\mathbold{w}_{j}^\top\mathbold{X}_{j}^\top\mathbold{X}_{N-1}
    \end{bmatrix} \label{non-linear-goal}
\end{aligned}
\end{equation}

\noindent Each term in (\ref{non-linear-goal}) has the similar format to the term in (\ref{q-goal}) except for $\mathbold{w}_{j}^\top \mathbold{X}_{j}^\top \mathbold{X}_{i}$ in \eqref{non-linear-goal} has coefficient $\frac{1}{4}$ and $\mathbold{y}^\top \mathbold{X}_{i}$ in \eqref{q-goal} becomes $(\mathbold{y}-\frac{1}{2})^\top \mathbold{X}_{i}$ in \eqref{non-linear-goal}. We can modify the protocol to compute the gradients for non-linear models without exposing labels $\mathbold{y}$ and $\mathbold{X}_{i}$. First, when the aggregator constructs function vector $\mathbold{c}$, instead of using the original weights, we multiply the weights with $\frac{1}{4}$ element-wise and use the modified weights to construct $\mathbold{c}$. Moreover, the active party, instead of sending the ciphertext of $\mathbold{y}$, now sends the ciphertext of $\mathbold{y}-\frac{1}{2}$. After decrypting and concatenating all the elements in \eqref{non-linear-goal}, the aggregator multiplies the concatenated results with $\frac{1}{S}$ to obtain the gradients. Other procedures remain same as the procedure for the linear regression models.